\documentclass[12pt]{article}%
\usepackage{amsmath}%
\usepackage{amsfonts}%
\usepackage{amssymb}%
\usepackage{graphicx}
\usepackage{tikz}      
\usepackage{multicol}  
\usepackage{chngpage}  
\usepackage{rotating}
\begin{document}
\begin{titlepage}
\title{The Value of A Statistical Life \\ in Absence of Panel Data: What can we do?}
\author{Andr\'es Riquelme \\ Marcela Parada}
\date{March 19, 2012}
\maketitle
\begin{center}
\textbf{Work in progress, do not cite.}
\end{center}
\thispagestyle{empty}
\vskip40pt
{\linespread{1} \begin{abstract}
In this paper I show how reliable estimates of the Value of a Statistical Life (VSL) can be obtained using cross sectional data using Garen's instrumental variable (IV) approach.  The increase in the range confidence intervals due to the IV setup can be reduced by a factor of 3 by using a proxy to risk attitude.  In order state the ``precision'' of the cross sectional VSL estimates I estimate the VSL using Chilean panel data and use them as benchmark for different cross sectional specifications.  The use of the proxy eliminates need for using hard-to-find instruments for the job risk level and narrows the confidence intervals for the workers in the Chilean labor market for the year 2009.
\end{abstract}
}
\end{titlepage}

\newpage
\section{Introduction}
In the last years the Value of a Statistical Life (VSL) has won an important place as a tool for public policy decision and welfare measure (Jena et al, 2003)\cite{j_2003}.  As consequence, there is an increasing concern on the reliability of the VSL estimates due to the disparate results and the sometimes excessively large confidence intervals found in the literature.  Several methods have been proposed to get reasonable and stable estimates, and the most successful ones involve the use of panel data sets (Knieser et al, 2012)\cite{k_2012}.  This arises one important concern: can we still say anything about the VSL in absence of panel data?  This concern becomes particularly important in developing economies, where there are neither VSL estimates or available panel data sets.

By using data from the Chilean labor market I show how the VSL can be accurately estimated from cross sectional data at a cost of a wide confidence interval by using Garen's (1988)\cite{g_1988} instrumental variable (IV) approach.  I also show that the range of this confidence interval can be reduced by a factor of 3 by using a proxy to risk attitude.  In order state the ``precision'' of the cross sectional VSL estimates I use panel data estimates as benchmark.

The innovation respect to the previous literature is the use of a variable that measures directly the willingness to take risks, which avoids the need for the use of instruments and at the same time reduces the variance of the estimates.

\section{Literature Review}
\label{sec:literature}

In addition to the traditional econometric problems in the estimation of wage equations (such as selection bias and omitted variables), the estimation of the Value of a Statistical Life using the hedonic wages approach adds the issue of the measurement error in the risk levels (fatal and non-fatal).  Authors have tried to account for this issue, especially when it is correlated with the covariates in the wage equations (Black and Kniesner, 2003)\cite{b_2003}, when the safety risks levels are difficult to quantify (Ashenfelter, 2006)\cite{a_2006} or when the non fatal measure is not as accurate as the fatal measurement (Siebert and Wei, 1994)\cite{s_1994}.

Even if we have a good measurement of the risk level, usually they are presented in an aggregate fashion, most of the time grouped by economic industry group.  If the aggregation is done in $j$ sectors, then the risk measurement variable will take only $j$ possible values.  This implies that we will end up with zero heterogeneity between groups and that the estimates will have low explanatory power.  Additional to this, we expect that in most of the cases the individuals will supply their work hours according to the perceived risk and not the actual risk level, which may be unknown. Viscusi (1979)\cite{v_1979} proposes that perceived risk is correlated with the effective risk rates, but usually researchers cannot test that hypothesis because very detailed data is necessary in order to have reliable estimators.  Timmins and Murdock (2007)\cite{t_2007} propose a method to alleviate such need for detail using data from other firms in the same economic sector as instruments for risk level perception. In this case, the risk level measure will be a good covariate for estimating the labor demand function but not for the labor supply function.

Another concern when the wage-risk trade-off is estimated is the self-selection problem.  It arises in two ways: First, workers with low risk aversion will choose risky jobs not necessarily because of the higher wage, but for the utility they obtain from risk, which implies that wages and risk levels are simultaneously determined (Garen, 1988)\cite{g_1988}.  Second, the omission of unobserved productivity lowers the VSL estimates.  According to Hwang et al. (1992)\cite{h_1992}, the addition of productivity proxies to the estimation of wage equations tends to increase the estimators of the VSL up to ten times, whereas Garen (1988)\cite{g_1988} suggested that ``coolheadness'' should explain differences in productivity: a cool-headed worker may be more productive that the average worker in a risky job, but non different than the average worker in a safe work.

The best solution the self selection problem is the use of panel data to difference out any unobservable ability correlated with occupation selection and risk preferences (Knieser et all, 2012)\cite{k_2012}.  Unfortunately panel data is not allways available.  A feasible solution using cross sectional data is the use of an appropriate instrumental variable (Reiersol, 1941)\cite{r_1941}, however there is wide consensus in the difficulty to find one.  Also, uncomfortably wide confidence intervals are obtained using weak instruments (Kochi, 2007)\cite{k_2007}.  

The most of the studies have concentrated in to obtain VSL estimates for developed countries. Since there are differences in the VSL estimates among developed and underdeveloped countries and am using data from one of the last, I will provide a brief discussion of the differences in the section \ref{sec:underdeveloped}.

\section{VSL Variations Between Countries}
\label{sec:underdeveloped}

Among the few studies with results for developed countries, Miller (2000)\cite{m_2000} is perhaps the most comprehensive.  Miller analyzes 68 studies and extrapolate the results to 49 countries using the Gross Domestic Product as predictor.  Even though the indirectly obtained values cannot be directly compared with those estimated using hedonic wages, they provide some insights about how the relative magnitudes of the estimates should be among countries.

As one can expect, the estimates for developed countries are higher than the estimates for developing and underdeveloped ones and some explanations can be provided.  First, the labor markets in underdeveloped countries are not properly developed, and then the market wages do not incorporate properly the risk levels, yielding small risk premiums.  Second, safety standards are less restrictive in those countries, allowing the employers to provide low quality safety equipment and facilities.  This problem is worse in the absence of a job contract, in which case the employer is not responsible for providing any safety measures and workers must accept a lower wage to face a reduction in the work risk.  Finally, the results for underdeveloped countries usually are expressed in U.S. dollars, but not corrected by parity of purchasing power.  

Whichever the cause, the difference is notorious if we compare the results reported by Miller (2000)\cite{m_2000} for North America and European Union of 2.3 to 5 million dollars of 2009 with the estimates for Chile that ranges form 0.8 to 1.3 million dollars.  However, since Miller's results are based in extrapolations of VSL from developed countries based on GDP, the results are not comparable and the estimates for underdeveloped countries are lower just because they have lower GDP per capita.  The only VSL result obtained from labor market data for estimate for Chile is about 4.6 million dollars (Parada et al, 2012)\cite{p_2012}.

\section{Econometric Framework}
\label{sec:econometric}

For the panel estimates I used the following hedonic log-wage equation:
\begin{equation} \label{eq:logwage}
\ln w_{ijt} = \alpha_i + \alpha_1 \pi_{jt} + X_{it} \beta + u_{ijt} \quad ,
\end{equation}
where
$\ln w_{ijt}$ is the log of the monthly wage of worker $i$ $(i=1,\dots ,N)$,
in the industry $j$ $(j=1, \dots , J)$,
at time $t$ $(t= 1, \dots ,T)$, 
$\alpha_i$ is an individual constant term,
$\pi_{jt}$ is the industry specific risk rate,
$X_{it}$ are demographics controls: education, age, age squared, monthly hours worked and dummies for public sector, union status and contract status.  $u_{ijt}$ is a normally distributed iid error term.  I tried different specifications using time controls also, but they were not significant.

The cross sectional estimates are the one-period version of the log-wage equation.  From equation (\ref{eq:logwage}) we estimate the VSL by using the formula
\begin{equation} \label{eq:vsl}
\widehat{VSL}=\left[ \left(\frac{\partial \hat{w}}{\partial \pi } = \hat{\alpha}_1 \times w \right) \times h \times 10,000 \right],
\end{equation}
were $h$ are the annual hours of work, $w$ is the average wage (Note that the fatal risk is per 10,000 worker.)


\section{Data}
\label{sec:data}
I have used two sources of information: the Chilean Safety Association Yearly Reports, which are the only public statistics about workplace fatal and non fatal accidents, and the Chilean Social Protection Survey, waves 2004, 2006 and 2009, which is a rich data set of approximately 20,000 Chilean people in the labor market.  It contains information about labor history, retirement funds, education, health, training, assets and household composition.  The survey is modular: individual, labor history and household information is presented in separated hierarchical data files.  In order to have comparable estimates I used male and female over 19 years old that worked for salary pay in the year of the survey, non permanently disabled and without missing values on the covariates{\linespread{1}\footnote{Perhaps for comparison with the bulk of results for developed economies it would be better to estimate only for male workers, but the fatality risk variable includes both genders, so separation is inappropriate.  Also, the unique VSL for Chile includes both male and female, so I did not restrict the sample.}}.  

The description of the workplace fatal risks by industry group and year are presented in table \ref{tab:risk}.  As shown in the table, the aggregation of the information is by one digit SIC, so the variable \emph{risk} is constant over groups.  

The descriptive statistics of all the variables for the econometric exercise are presented in table \ref{tab:descriptive} for the wave of 2009. 

 The common variables for the Mincer's specification are presented along with a dummy for the existence of job contract.  This is an important variable usually not presented in studies for developed countries, but that makes a big difference in underdeveloped ones as explained in section \ref{sec:underdeveloped}. The instruments for risk are marital status, spouse years of schooling, dummies for spouse illness and job and the number of children under 6 years old.  The variable \emph{willingness to take risks} is the used proxy for risk preferences and consists of a likert-type variable ranging from 0 to 10, 0 being the lesser willingness to take risks.  Under the assumption that risk preferences do affect job choice and  wages, the inclusion this variable will capture the individual heterogeneity eliminating the need for an instrument. 

\section{Results}
\label{sec:results}

The benchmark estimates for the equation \ref{eq:vsl} are presented in table \ref{tab:panel}.  As expected, I get different results from the different panel specifications, ranging between 1.4 and 7.4 million dollars.  The pooled OLS and the fixed estimators are in general expected to be either downwards or upward bias due to unobservable heterogeneity.  If the unobserved variables that capture heterogeneity are time invariant, then the within estimator or the first difference estimator should be the most appropriated.  The point estimators for this two models are similar, but the first differences estimator is not significant.  This is not surprising in presence of small unbalanced panel.  The minimum requirement for identification of a balanced model is two sample periods and in this case there are just three.  As pointed by Jimenez-Martin (1988)\cite{j_1998}, unbalanced panel exhibit worse testing performance than balanced one for a fixed number of individuals.  If the unobserved variables are not time invariant, then the random effects will be more appropriated.  Note that both random effects estimators are significant and lies between the lowest and highest estimates, so am using them as benchmark.  Although the random effects model assumes that unobserved heterogeneity is uncorrelated with the covariates in the model, which may not be true, it still reduce the VSL compared with those who are not controlling for it (Knieser et al, 2012)\cite{k_2012}.

Assuming that the ``right'' estimate for the VSL is around 3.4 and 3.9 million dollars, the next step is to investigate if similar results can be obtained from cross sectional data.  

There results of the estimation of the cross sectional version of the equation \ref{eq:vsl} are presented in the table \ref{tab:results}.  In the first column I show the base estimates for the log-wage equation using the simple linear approach.  The right hand side variables are a constant, years of schooling, age, age squared, worked hours, three dummies for public sector workers, contract and union membership and the fatal risk.  The estimated VSL is us\$715,036 with a range of the 95\% confidence interval of us\$133,159. This estimate is expected to biased because is not correcting for unobserved heterogeneity and it is outside the confidence intervals of the benchmark estimates.  The effect of the bias is undetermined and depends on how the unobserved variables are correlated with the wage and fatality rates.

The result of the use of the IV estimation using Garen's (1988) approach is presented in the column (2).  This specification has the same covariates as those in column (1), but the risk is instrumented using marital status, number children under 6 years old, schooling of the spouse and dummies for illness of the spouse and  work of the spouse. As expected, the estimate VSL increased to us\$3,654,338, but at a cost of an increase in the range of the 95\% confidence interval, which is now us\$819,741.  In spite of the increase in the confidence interval, this estimate is more apropriate than that obtained in the first setup because it is controlling for risk preferences and it is statistically not different from the benchmark estimates and has lower variance.  This is mild evidence in favor of the goodness of the instruments. 

Overall, the instruments are good predictors of risk and satisfy the exclusion condition under the Anderson-Rubin (1949)\cite{a_1949} test setup, but they do not under the Fractionally Resampled Anderson-Rubin setup (Berkowitz et al, 2012)\cite{b_2012}, so one or some of them may be weak. 

The final specification is presented in the column (3). This estimation include the same covariates than in the column (1) and adds the proxy for the risk preferences.  The estimate for the fatal risk coefficient is not statistically different from that obtained from the IV estimation, but its standard error is smaller.  This allow a good reduction in the width of the 95\% confidence interval to us\$265,153.  This is only a third of the range estimated by using the IV approach and is evidence supporting the distortion of the use of inappropriate instruments (Kochi, 2007)\cite{k_2007}.

To put this estimates in context, they are statistically lower than the reported by Knieser et al. (2012)\cite{k_2012}, which ranges the VSL for USA between 5 to 12 million dollars and also inferior to the ones reported by Parada et al (2012)\cite{p_2012} for Chile, but since there is not an interval for the last ones, equality cannot be either accepted or rejected.

\section{Conclusion}
In this paper I have shown that it is still possible to get reliable VSL estimates from cross sectional data.  This is crucial in developing economies, where the available data is scarce or non-existent and the cost of a panel-type surveying process is sometimes too high.  In such cases, dimension reductions are very welcome as ways to reduce surveying costs and increase the quality of statistical inference by using better covariates.  I also provided evidence that the instruments proposed by Garen (1988)\cite{g_1988} may be weak and better instrument are required to use this approach. 

This results provides evidence that that the use of a variable that directly measure the attitude toward risks can eliminate the need of the use of instruments to correct for risk and wage endogeneity and reduce the range of the VSL confidence intervals.

It is important to note that this approach is not eliminating the measurement error in the job risk variable and it is not controlling for latent heterogeneity in productivity, but still provides a good way to increase the accuracy of the VSL estimates.  One concern is that the proxy for risk may not be capturing the attitude towards labor risk.  The question \textit{How willing are you to take risks}? is too broad, and people can be thinking different risk sources when answering (like financial risks or risky outdoors activities.) This arises a measurement error problem for the proxy, and the parameters may be biased toward zero, but since the point estimates under the IV and under the linear setup are the same, there is no evidence that the bias is important.

\linespread{1.1}

\newpage
\section{Appendix: Tables}
\begin{table}[htp]
\center 
\caption {\label{tab:risk}Fatal Risk by Industry Group.}
\begin{tabular}{l r@{}l r@{}l r@{}l } \hline \hline
Group                                   & \multicolumn{2}{c}{ \ 2009} & \multicolumn{2}{c}{ \ 2006}  & \multicolumn{2}{l}{ \ 2004} \\ \hline
Agriculture, Forestry and Fisheries      &   19&.6  & 11&.4  &    14&.2  \\
Mining                                   &   22&.0  & 37&.2  &     8&.8  \\
Manufacturing                            &    6&.4  &  8&.2  &    10&.4  \\
Electricity, Gas and Water               &   34&.2  &  0&.0  &   253&.4  \\
Construction                             &   22&.6  & 25&.3  &    23&.0  \\
Trade                                    &    5&.2  &  9&.1  &     2&.1  \\
Transportation, Storage \& Communication &   29&.1  & 35&.4  &    23&.3  \\
Services                                 &    7&.0  &  7&.9  &     3&.6  \\ \cline{2-7}
Weighted Average                         &    2&.1  &  1&.4  &     1&.5  \\ 
Number of observations                   & 4,9&53   &  5,9&51  &  6,0&47    \\ \hline \hline
\multicolumn{7}{l}{\footnotesize $a$: Fatal risk in 1/10,000 ratio.}\\
\multicolumn{7}{l}{\footnotesize Source: Chilean Safety Association.}
\end{tabular}
\end{table}

\begin{table}[p]
\center
\caption {\label{tab:descriptive}Descriptive Statistics of Selected Variables.}
\begin{tabular}{lcccc} \hline \hline
                      & Min   & Max       & Mean   & Std. Dev.\\ \cline{2-5}
Monthly wage$^a$         & 2.144 & 26,802 & 573.73 & 872.055\\
Log Wage              & 0.763 & 10.196    & 6.01   & 0.783\\
Monthly hours worked  & 2     & 126       & 44.91  & 13.160\\
Years of schooling    & 0     & 25        & 10.74  & 4.202\\
Age                   & 19    & 84        & 44.55  & 11.376\\
Age Squared           & 361   & 7,056     & 2,114  & 1,064\\
Gender                & 0     & 1         & 0.60   & 0.490\\
Public sector worker  & 0     & 1         & 0.13   & 0.336\\
Have a contract       & 0     & 1         & 0.67   & 0.47\\
Union affiliation     & 0     & 1         & 0.17   & 0.38\\
\textit{Instruments for Risk}& &          &        &    \\ 
\ \ \ \ Marital Status        & 0     & 1         & 0.63    & 0.482\\
\ \ \ \ Spouses's Schooling   & 0     & 24        & 5.89    & 5.755\\
\ \ \ \ Spouses's Illness     & 0     & 1         & 0.58    & 0.493 \\
\ \ \ \ Spouses Works         & 0     & 1         & 0.29    & 0.453\\
\ \ \ \ Children under 6      & 0     & 2         & 0.13   & 0.355 \\
\textit{Risk Measures}& &          &         &    \\
\ \ \ \ Fatal Risk (1/10,000) &2.635  &28.891     & 8.848    &8.390\\ 
\ \ \ \ Willingness to take risks & 0 & 10        & 5.6     & 3.1\\ \cline{1-5}
Number of observations   & 6,059 &           &  & \\ \hline \hline
\multicolumn{5}{l}{\footnotesize $a$: Expressed in US dollars of 2009. }\\
\multicolumn{5}{l}{\footnotesize Source: Social Protection Survey 2009, Chile. }\\
\end{tabular}
\end{table}

\begin{sidewaystable}[p]
\caption {\label{tab:panel}Panel Estimates of the Wage-Fatal Risk Trade-off.}
\begin{tabular}{lcccccc} \hline \hline
                               & Pooled OLS & Between      & Within    &First      & Random   &\ \ Random    \\
                               & Estimator  &Estimator     &Estimator  &Difference &  Effects &\ \ Effects   \\
                               &   (1)      &  (2)         & (3)       & (4)       & GLS (5)  &\ \ MLE (6)   \\ \cline{2-7}
Risk Coefficient $\times$ 1000 &  2.6918    & 3.9378       & 0.7485    &0.8092     &1.8284    &\ \ 2.0791    \\
                               & (0.0005)    &(0.0039)      & (0.0007)   &(0.0011)   &(0.0018) &\ \ (0.0005) \\
$\widehat{VSL}$  (millions)    & 5.1        & 7.4          &1.4        &1.5        &3.4       &\ \ 3.9       \\
95\% Confidence Interval       &[3.4|6.8]   & [3.8 | 11.0] &[0.2 | 2.6]&[-2.6|5.6] &[2.0|4.9] &\ \ [2.0|5.7] \\
Variation Range                & 1.7        &   3.6        &   1.2     & 4.1       & 1.4      &\ \ 1.8\\  \cline{1-7}
Number of Observations         & 13,387     &  6,913       &  13,387   &6,577      & 13,387   &\ \ 31,387\\ \hline \hline
\end{tabular}
\footnotesize
Dependent variable in all estimations is log wage.  All the models controls for education, age, age squared, hours of work and dummies for public sector workers, contract existence, union affiliation and time. Robust standard errors are in parentheses.

\end{sidewaystable}

\begin{sidewaystable}[p]
\caption {\label{tab:results}Estimates of the Wage-Fatal Risk Trade-off, year 2009.}
\begin{tabular}{lccc} \hline \hline
                             \ \ \  & Linear Wage        \ \ \   & IV Estimation         \ \ \  & Linear Wage  \\
                             \ \ \  &  Equation (1)      \ \ \   &   (2)                 \ \ \  &  Equation (3) \\ \cline{2-4}
Risk Coefficient $\times$ 10 \ \ \  &  1.0386            \ \ \   & 5.3078                \ \ \  & 5.4602   \\
                             \ \ \  & (0.0010)           \ \ \   &(0.0058)               \ \ \  & (0.0020) \\
$\widehat{VSL}$              \ \ \  & 715,036            \ \ \   & 3,654,338             \ \ \  &3,759,271 \\
95\% Confidence Interval     \ \ \  &[580,204 | 849,869] \ \ \   &[2,865,950 | 4,442,726]\ \ \  &[3,494,118 | 4,024,424] \\
Variation Range              \ \ \  & 134,833            \ \ \   &   788,388             \ \ \  &  265,153 \\  \cline{1-2}
Number of Observations              & 6,059                      &                             & \\ \hline \hline
\end{tabular}
\footnotesize
Dependent variable in all estimations is log wage.  All models controls for education, age, age squared, hours of work and dummies for public sector workers, contract presence and union affiliation. Model (2) instrumentalize the fatal risk using marital status, number children under 6 years old, schooling of the spouse and dummies for illness of the spouse and  work of the spouse as suggested in Garen (1988).  Model (3) uses willingness to take risk as proxy for risk preferences.  Heteroskedasticity-robust standard errors are in parentheses.

\end{sidewaystable}


\end{document}